\documentclass[twocolumn,showpacs,preprintnumbers,amsmath,amssymb,superscriptaddress,pra]{revtex4}

\usepackage{graphicx}
\usepackage{bm}
\usepackage{braket}
\usepackage{comment}
\usepackage{amsmath}
\usepackage{amssymb}

\newcommand{\myeqnref}[1]{Eq.~\eqref{#1}}
\newtheorem{claim}{Claim}

\newcommand{\figlength}{3.5in}

\begin{document}


\title{Phase-Remapping Attack in Practical Quantum Key Distribution Systems}

\author{Chi-Hang Fred Fung}%
 \email{cffung@comm.utoronto.ca}
\affiliation{%
Center for Quantum Information and Quantum Control,\\
Department of Electrical \& Computer Engineering and Department of Physics,\\
University of Toronto, Toronto,  Ontario, Canada\\
}%

\author{Bing Qi}%
 \email{bqi@physics.utoronto.ca}
\affiliation{%
Center for Quantum Information and Quantum Control,\\
Department of Electrical \& Computer Engineering and Department of Physics,\\
University of Toronto, Toronto,  Ontario, Canada\\
}%

\author{Kiyoshi Tamaki}
 \email{tamaki@will.brl.ntt.co.jp}
\affiliation{%
NTT Basic Research Laboratories, NTT corporation,\\
3-1,Morinosato Wakamiya Atsugi-Shi, Kanagawa, 243-0198; \\
CREST, JST Agency, 4-1-8 Honcho, Kawaguchi, Saitama, 332-0012, Japan \\
}%

\author{Hoi-Kwong Lo}
 \email{hklo@comm.utoronto.ca}
\affiliation{%
Center for Quantum Information and Quantum Control,\\
Department of Electrical \& Computer Engineering and Department of Physics,\\
University of Toronto, Toronto,  Ontario, Canada\\
}%



\begin{abstract}
Quantum key distribution (QKD) can be used to generate secret keys
between two distant parties. Even though QKD has been proven
unconditionally secure against eavesdroppers with unlimited
computation power, practical implementations of QKD may contain
loopholes that may lead to the generated secret keys being
compromised. In this paper, we propose a phase-remapping attack
targeting two practical bidirectional QKD systems (the ``plug \&
play'' system and the Sagnac system). We showed that if the users of
the systems are unaware of our attack, the final key shared between
them can be compromised in some situations.
Specifically, we showed that, in the case of the
Bennett-Brassard 1984 (BB84) protocol with ideal single-photon
sources, when the quantum bit error rate (QBER) is between $14.6\%$
and $20\%$, our attack renders the final key insecure, whereas the
same range of QBER values has been proved secure if the two users
are unaware of our attack; also, we demonstrated three situations
with realistic devices where positive key rates are obtained without
the consideration of Trojan horse attacks but in fact no key can be
distilled.
We remark that our attack is feasible with only current technology.
Therefore, it is very important to be aware of our attack in order
to ensure absolute security.
In finding our attack, we minimize the QBER over individual measurements described by a general POVM, which has some similarity with the standard quantum state discrimination problem.
\end{abstract}


\maketitle

\section{Introduction}

One important practical application of quantum information is quantum key distribution (QKD) \cite{Bennett1984,Ekert1991,Gisin2002}, which generates secret keys between two distant parties, commonly known as Alice and Bob.
The advantage of QKD is that it has been proven {\it unconditionally secure} even when an eavesdropper, Eve, has unlimited computation power allowed by the law of quantum mechanics
\cite{Mayers2001,Biham2000,Lo1999,Shor2000,Inamori2005,Gottesman2004}.
On the other hand,
security proofs are only as good as their assumptions that
real-life QKD systems may not accomplish due to imperfections.
This may open up new attacks for Eve.
Moreover, given a combination
of imperfections, Eve may try to mix and pick the best (perhaps a combined)
eavesdropping strategy to maximize her chance of breaking a
QKD system.
It is thus important to construct a catalog of known attacks against
practical QKD systems.
Moreover, it is imperative to study specific defenses against proposed attacks.
Notice that implementations of defenses may open
up new security loopholes.
It is not enough to say that defense strategies
exist in principle.
One must also battle-test them thoroughly in experiments to see if they are of any good in practice.
We remark that the construction of
generally agreed theory of eavesdropping attacks and defenses in realistic
``plug-and-play'' systems
is, in fact, a five-year goal in the US funding agency ARDA's quantum
cryptography roadmap \cite{arda}.

Practical difficulties associated with phase and polarization instabilities over long-distance fiber have led to the development of
two bidirectional QKD structures:
the ``plug \& play'' auto-compensating QKD structure
\cite{Muller1997} and the Sagnac QKD structure
\cite{Nishioka2002,Qi2006a}. In both cases, one of the legitimate
users, Bob, sends strong laser pulses to the other user, Alice.
Alice encodes her information on the phase of the strong pulse,
attenuates it to single photon level, and then sends it back to Bob.
Because Alice allows signals to go in and go out of her device, this
opens a potential backdoor for Eve to launch various Trojan horse
attacks, which are any attacks that involve more than just passive
attacks.
Trojan horse attacks performed by sending probe signals into Alice's and Bob's equipments have been analyzed in \cite{Gisin2005};
Trojan horse attacks exploiting the detector efficiency mismatch have been analyzed in \cite{Makarov2006} and also by us \cite{Qi2007a}.
In this paper, we propose a specific type of Trojan horse attack, which we call the phase-remapping attack aiming at bidirectional QKD system using phase coding.
We show that, when Alice and Bob are unaware of our attack, the final key shared between them can be compromised in some situations.
Also, our attack is feasible with only current technology.
Therefore, it is very important for Alice and Bob to be aware of our attack when using the ``plug \& play'' QKD systems or the Sagnac QKD systems and to correctly identify which situations are secure and which are not.

In the following, we first describe in Sec.~\ref{sec-Sagnac}
and Sec.~\ref{sec-pnp} how phase remapping is performed in the two
QKD systems implementing the Bennett-Brassard 1984 (BB84) protocol
\cite{Bennett1984}, and then we illustrate situations in which the
final keys can be compromised, both in the
perfect-single-photon-source case and in the
weak-coherent-state-source case.
For the perfect-single-photon-source case (Sec.~\ref{sec-singlephoton}),
we aim to find the smallest quantum bit error rate (QBER) under the phase-remapping attack and show that
it is lower than the known QBER threshold under which secret keys can be distilled when Trojan horse attacks are not taken into account.
We formulate our problem as minimizing the QBER over an individual measurement described by a general POVM.
For the weak-coherent-state-source case (Sec.~\ref{sec-coherent}),
we demonstrates three specific eavesdropping strategies with the
phase-remapping attack (two of them are also combined with the
fake signals attack \cite{Makarov2006} which
exploits detection efficiency mismatch between two detectors) that lead Alice and Bob to
wrongly believe that they can distill secret keys at positive rates
but in fact no secret key can be generated. We finally conclude in
Sec.~\ref{sec-conclusions}.

\section{Phase-remapping attack in Sagnac QKD systems\label{sec-Sagnac}}

\begin{figure}
\centering
\includegraphics[width=\figlength]{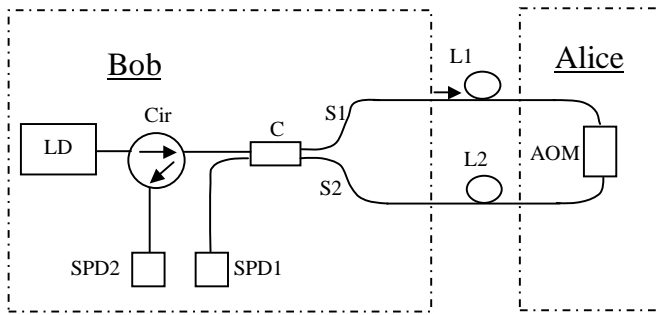}
\caption{\label{fig-Sagnac}
Schematic diagram of the Sagnac QKD system employing AOM-based phase modulator:
LD - pulsed laser diode; Cir - circulator; C - 2x2 coupler; SPD1, SPD2 - Single Photon detector
}
\end{figure}

The basic structure of the Sagnac QKD system \cite{Qi2006a} is shown in Fig.\ref{fig-Sagnac}.
Here, to simplify our discussion, we neglect Bob's phase modulator.
Note that we use an acoustic-optic modulator (AOM) as a phase modulator on Alice's side.
The input laser pulse is split by the fiber coupler into $S_1$ and $S_2$, which go through the fiber loop clockwise and counterclockwise, respectively.
Note that the AOM is placed in the fiber loop asymmetrically, with fiber lengths $L_1$ and $L_2$ on the two sides.
For the first order diffracted light, the AOM introduces a frequency shift equal to its driving frequency (due to Doppler effect).
The phase of the diffracted light is also shifted by an amount which is equal to the phase of the acoustic wave at the time of diffraction \cite{Stefanov2003}.
$S_2$ and $S_1$ arrive at the AOM at different times with the time difference $t_2-t_1=n(L_2-L_1)/C=n \Delta L/C$.
Here, $n$ is refractive index of optical fiber and $C$ is the speed of light in vacuum.
The phase difference between $S_1$ and $S_2$ after they go through the fiber loop is
\begin{equation}
\label{eqn-phasediff}
\Delta \phi = \phi(t_2)-\phi(t_1) = 2 \pi f (t_2-t_1) = 2 \pi n \Delta L f / C .
\end{equation}
By modulating the AOM's driving frequency $f$, the relative phase between $S_1$ and $S_2$ can be modulated.
This is the basic mechanism of our AOM-based phase modulator.

In standard BB84, Alice can encode phase information $\{0, \pi/2,
\pi, 3\pi/2\}$ by modulating the AOM with frequency $\{f_0,
f_0+\Delta f, f_0+2\Delta f, f_0+3\Delta f\}$. From
\myeqnref{eqn-phasediff}, the phase difference depends on both the
AOM frequency $f$ and the fiber length difference $\Delta L$. So, in
principle, Eve can build a device similar to Bob's one except with
different fiber length and launch an ``intercept-and-resend''
attack.

\begin{figure}
\centering
\includegraphics[width=\figlength]{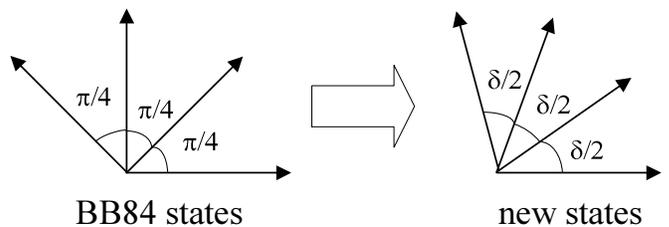}
\caption{\label{fig-fourstates}
The phase difference between the four states sent by Alice is changed by Eve to $\delta$.
In standard BB84, $\delta=\pi/2$.
(Note that the states are drawn so that orthogonal states are $\pi/2$ apart in the diagram but are $\pi$ apart in the actual phases.)
}
\end{figure}

Suppose Eve uses her device to send laser pulses to Alice.
Unaware that the pulses come from Eve, Alice shifts the light frequency by one of the values $\{f_0, f_0+\Delta f, f_0+2\Delta f, f_0+3\Delta f\}$.
By choosing a suitable fiber length difference $L_2-L_1$, Eve can re-map the encoded phase information from $\{0, \pi/2, \pi, 3\pi/2\}$ to $\{0, \delta, 2\delta, 3\delta\}$, where $\delta$ is under Eve's control.
This is illustrated in Fig.~\ref{fig-fourstates}.

\section{Phase-remapping attack in ``plug \& play'' systems\label{sec-pnp}}

\begin{figure}
\centering
\includegraphics[width=\figlength]{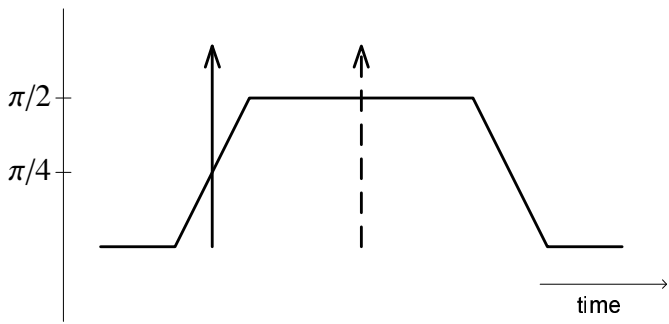}
\caption{\label{fig-risingedge}
The dashed line is the original signal pulse intended to be modulated at the middle of the phase modulator's response to have a phase of $\pi/2$.
Eve time shifts the pulse to the one in solid line.
This pulse now arrives at the middle of the rising edge and acquires a phase of $\pi/4$ instead.
}
\end{figure}

In a ``plug \& play'' QKD system \cite{Muller1997}, the information is encoded on the relative phase between a signal pulse and a reference pulse.
The phase modulator inside Alice is supposed to be activated in such a way that only the signal pulse is modulated while the reference pulse is not.
Unfortunately, in current QKD systems, Alice does not monitor the arrival times of the two pulses.
Instead, she just uses one of them as the trigger signal to determine when she should activate her phase modulator.
In this case, Eve can time-shift the signal pulse so that it will arrive at the phase modulator on its rising or falling edge and thus will be partially modulated (see Fig.~\ref{fig-risingedge}).
(The $LiNbO_3$ waveguide-based phase modulators used in current QKD systems have rise times ranging from $100ps$ to $1ns$).
Therefore, the relative phase between the signal pulse and reference pulse will be smaller than what it is supposed to be.
In principle, by carefully controlling the amount of time shift, Eve can re-map the encoded  phase information from  $\{0, \pi/2, \pi, 3\pi/2\}$ to $\{0, \delta, 2\delta, 3\delta\}$, where $\delta \in [0,\pi/2]$.

\section{Upper bound on QBER of phase-remapping attack with a perfect single-photon source\label{sec-singlephoton}}

We have described the possibility of Eve changing the phase difference $\delta$ between the states sent by Alice in two practical QKD systems.
The important question is: is this ability of Eve harmful to Alice and Bob in any way?
As we show in this section, Eve can use this ability to compromise the final key shared between Alice and Bob under some situations in the perfect-single-photon-source case.
We show this by considering Eve launching a specific intercept-and-resend attack that is optimized for the phase difference $\delta$ that she has chosen for Alice's states.
Note that any intercept-and-resend attack completely breaks the security of any QKD protocol, meaning that Alice and Bob cannot establish a secret key of any length \cite{Curty2004}.
Thus, we want to show that our intercept-and-resend attack leads to situations that Alice and Bob (wrongly) believe that they can generate a secret key.
The quantum bit error rate (QBER) is often used as a measure to judge whether a secret key can be generated in a QKD experiment.
The QBER can be obtained by Alice and Bob in a QKD experiment by publicly testing the error rates in a random subset of the transmitted bits.
They use the QBER to determine the amount of eavesdropping on the channel and whether to proceed with the key generation process.
Therefore, we want to show that our intercept-and-resend attack causes a quantum bit error rate (QBER) that is {\it lower} than what is tolerable without any Trojan horse attacks.
In this case, there is a range of QBER's that is secure without any Trojan horse attacks but is now insecure with our Trojan horse attack.
If Alice and Bob are unaware of our Trojan horse attack and treat these situations as secure, then their final secret key is compromised and Eve has some information on it.
In the following, we first consider an intercept-and-resend attack preceded by the phase-remapping operation.
In this attack, Eve's measurement is optimized and the resent states are the BB84 states.
We then consider three extensions to the attack strategy by optimizing over the resent states and/or combining the phase remapping attack with the fake signals attack \cite{Makarov2006}.
In all cases, we show that the final key can be compromised if no Trojan horse attack is considered.

\subsection{A simple intercept-and-resend attack with phase remapping}

We consider the BB84 protocol with a perfect single-photon source and detectors.
Note that any QBER lower than $20\%$
is tolerable in BB84 without any Trojan horse attacks
\cite{Gottesman2003,Chau2002,Ranade2006}, meaning that a secret key
can be distilled. Thus, we aim to construct an intercept-and-resend
attack that produces a QBER lower than this. The
intercept-and-resend attack we consider here is similar to the one
considered earlier by us \cite{Fung2006}. Here, we optimize the
attack to the phase difference between Alice's states, $\delta$,
which is set by Eve.


The four states sent by Alice have phases $0, \delta, 2\delta,$ and  $3\delta$, where the phase offset is set to be zero for simplicity and without loss of generality.
We assume that Eve uses the same detection scheme as Bob does.
Thus, for a state with phase $\theta$, Eve detects the bit values ``0'' and ``1'' with probabilities $\cos^2(\frac{\theta}{2})$ and $\sin^2(\frac{\theta}{2})$, respectively.
To facilitate the analysis, we denote Alice's four states as
\begin{eqnarray}
\label{eqn-phaseremappedstates}
\ket{\tilde{\varphi}_k} &=& \cos\left(\frac{k\delta}{2}\right) \ket{0_z} + \sin\left(\frac{k\delta}{2}\right) \ket{1_z}
\end{eqnarray}
where $k=0,\ldots,3$ are the indices for the four states, and
$\ket{j_z}, j=0,1$ are the eigenstates of the $Z$ component of the
Pauli matrix representing the bit values ``$j$''. Similarly,
$\ket{j_x}=(\ket{0_z}+ (-1)^j\ket{1_z})/\sqrt{2}, j=0,1$ are the
eigenstates of the $X$ component of the Pauli matrix.
Here, $\ket{\tilde{\varphi}_0}$ and $\ket{\tilde{\varphi}_2}$ are ``0'' and ``1'' in one basis, whereas $\ket{\tilde{\varphi}_1}$ and $\ket{\tilde{\varphi}_3}$ are ``0'' and ``1'' in the other basis.
Note that the normal BB84 states have the phase difference $\delta=\pi/2$;
we denote the BB84 states as $\ket{\varphi_k}$.

We consider the following intercept-and-resend attack by Eve:
Eve captures the state sent by Alice, $\ket{\tilde{\varphi}_k}$, and perform a POVM measurement on it.
The POVM consists of five elements, $\{M_{\text{vac}},M_i:i=0,\ldots,3\}$, with $M_{vac}+\sum_{i=0}^3 M_i=\mathbf{I}$.
For the outcome corresponding to $M_{\text{vac}}$, Eve sends a vacuum state to Bob,
whereas, for outcome $i$, she sends the BB84 state $\ket{\varphi_i}$ to Bob.


\begin{figure}
\includegraphics[width=\figlength]{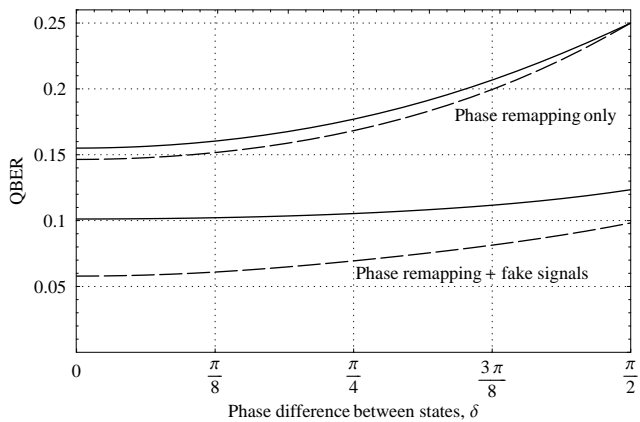}
\caption{\label{fig:upperbound_BB84} QBER upper bound of Trojan
horse attacks for BB84. The top two curves correspond to the
phase-remapping attack only whereas the bottom two cures correspond
to the combination of the phase-remapping attack and the
fake signals attack of Ref.~\cite{Makarov2006} (with efficiency
mismatch of $0.08$). The QBER of the two solid curves are obtained
by minimizing over the POVM measurement by Eve for each phase
difference $\delta$ and assuming a fixed state sent to Bob. The QBER
of the two dashed curves are obtained by minimizing over the POVM
measurement by Eve and the state sent to Bob for each phase
difference $\delta$. Note that the QBER values approach some minimum
values ($15.5\%$, $14.6\%$, $10.1\%$, and $5.79\%$) as the phase
difference between the states approaches zero.
}
\end{figure}

For a fixed phase difference $\delta$, we want to favor Eve by minimizing the QBER caused by this attack over the POVM elements.
This QBER minimization problem is similar to the quantum state discrimination problem \cite{Chefles2000}, where a given state is to be identified among a set of known states.
In our case, since the four states are not linearly independent, unambiguous discrimination (meaning error free) is not possible \cite{Chefles1998}.
In the standard ambiguous state discrimination problem, the total probability of incorrectly identifying the state $\sum_{i \neq j} \text{Tr}(M_i \ket{\tilde{\varphi}_j}\bra{\tilde{\varphi}_j}) /4$ is minimized subject to $\sum_{i=0}^3 M_i=\mathbf{I}$, where the division by four is due to Alice sending one of the four states with equal probabilities.
On the other hand, in our problem, the quantity to minimize is the QBER, which is the error rate on Bob's measured signals, not Eve's error probability.
We find the QBER as follows.
Consider $M_0$ first.
When $M_0$ occurs, Eve sends $\ket{{\varphi}_0}$ to Bob.
If Alice actually sent $\ket{\tilde{\varphi}_0}$, then there is no error.
However, if Alice actually sent $\ket{\tilde{\varphi}_2}$ and Bob uses the measurement basis $\{\ket{{\varphi}_0},\ket{{\varphi}_2}\}$ (only the cases that Alice and Bob use the same basis are considered), then Bob always gets an error and thus the QBER is $1$;
on the other hand, if Alice actually sent $\ket{\tilde{\varphi}_1}$ or $\ket{\tilde{\varphi}_3}$ and Bob uses the measurement basis $\{\ket{{\varphi}_1},\ket{{\varphi}_3}\}$, then the QBER is only $1/2$.
Therefore, the (unnormalized) QBER for the $M_0$ case is $[\frac{1}{2}\text{Tr}(M_0 \ket{\tilde{\varphi}_1}\bra{\tilde{\varphi}_1}) +\text{Tr}(M_0 \ket{\tilde{\varphi}_2}\bra{\tilde{\varphi}_2}) +\frac{1}{2}\text{Tr}(M_0 \ket{\tilde{\varphi}_3}\bra{\tilde{\varphi}_3}) ]/4$.
Comparing this with the total error probability of the state discrimination problem, we see that here different penalties are incurred for different incorrectly identified states.
To form the final QBER, we need to add the (unnormalized) QBER for the other $M_i$'s and normalize the sum
with the probability of Eve causing clicks on Bob's
detectors, giving us
\begin{eqnarray}
\label{eqn-QBER1} \text{QBER} &=& \frac{ \sum_{i=0}^{3}
\text{Tr}(M_i L_i) }{ \sum_{i=0}^{3} \text{Tr}(M_i B_i) },
\end{eqnarray}
where
\begin{eqnarray}
\label{eqn-QBER2}
L_i &=& \frac{1}{2} \ket{\tilde{\varphi}_{1+i}}\bra{\tilde{\varphi}_{1+i}}
 + \ket{\tilde{\varphi}_{2+i}}\bra{\tilde{\varphi}_{2+i}} \nonumber \\
&& +\frac{1}{2} \ket{\tilde{\varphi}_{3+i}}\bra{\tilde{\varphi}_{3+i}}, \\
\label{eqn-QBER3}
B_i &=& \sum_{k=0}^3
\ket{\tilde{\varphi}_{k}}\bra{\tilde{\varphi}_{k}}.
\end{eqnarray}
We minimize the QBER over positive $M_i$'s (see Appendix for detail).
Note that it is not necessary to impose the constraint $\sum_{i=0}^3 M_i \leq \mathbf{I}$, since any solution to this unconstrained problem can always be scaled down sufficiently to satisfy this constraint.
Also note that normalization of the QBER is necessary since we allow Eve to get an inconclusive result and send a vacuum state to Bob (i.e., we allow $M_{\text{vac}}$ to be non-zero).
This is in contrast to the standard ambiguous state discrimination problem where all results have to be conclusive.

In general, Eve's action is a solution to some optimization problem, minimizing some general penalty function.
The QBER and the total error probability in the standard state discrimination problem are two special cases of such general penalty functions.
In our Trojan horse attack problem, we use the QBER as the objective function since Alice and Bob can determine this value experimentally and use this value to estimate the amount of eavesdropping on the quantum channel.



\begin{figure}
\includegraphics[width=\figlength]{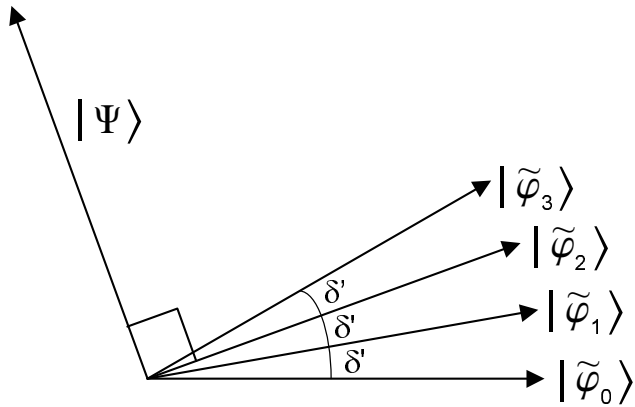}
\caption{\label{fig:smalldelta}
A suboptimal strategy for Eve.
She chooses $M_0=\ket{\Psi}\bra{\Psi}$ where $\ket{\Psi}$ is a state orthogonal to $\ket{\tilde{\varphi}_2}$.
This strategy causes a QBER of $16.7\%$.
}
\end{figure}

Figure~\ref{fig:upperbound_BB84} plots the smallest QBER induced by
this attack against the phase difference $\delta$ (top curve).
This curve is achieved by Eve resending only the states $\ket{0_z}$
and/or $\ket{1_x}$ to Bob.  Due to the symmetry in their
phase-remapped states $\ket{\tilde{\varphi}_0}$ and
$\ket{\tilde{\varphi}_3}$, the resultant QBER's are equal (see
Fig.~\ref{fig:smalldelta}). Also, it turns out that the QBER caused
by resending the states $\ket{1_z}$ or $\ket{0_x}$ is higher than
this curve in the range of $\delta$ shown in the figure. We
observe that this QBER curve approaches $15.5\%$ as the phase
difference $\delta$ approaches zero.
Note that there is a discontinuity at $\delta=0$.
When the phase difference is exactly zero, all four states sent by
Alice are exactly the same. Thus, Eve cannot learn anything about
Alice's bits. In this case, Eve can either send random states to Bob
(in which case the QBER is $\frac{1}{2}$) or send only vacuum states
to Bob (in which case the QBER is undefined since Bob did not have
any click). The source of this discontinuity is that we allow Eve to
get an inconclusive result and send a vacuum state to Bob (i.e.,
$M_{vac}\neq \mathbf{0}$).
Note that in practice, one may
restrict Eve's strategies by requiring a certain minimum detection
probability at Bob's side, meaning that Eve has to resend some
states to Bob with a minimum probability. As a consequence, Eve may
launch our attack only at phase differences $\delta$ larger than
some small finite value, in which case, the discontinuity at
$\delta=0$ is irrelevant.
In the standard state discrimination
problem, no inconclusive result is allowed and thus the error
probability approaches $1/2$ as $\delta$ approaches zero with no
discontinuity.

We can understand the behaviour of the top curve in Fig.~\ref{fig:upperbound_BB84} at small $\delta$ by considering a suboptimal intercept-and-resend strategy for Eve.
Let's consider that Eve is only interested in finding a good $M_0$ and assigns $M_1=M_2=M_3=0$.
Since $\ket{\tilde{\varphi}_2}$ causes the largest QBER of 1 (whereas $\ket{\tilde{\varphi}_1}$ and $\ket{\tilde{\varphi}_3}$ cause only $1/2$),
Eve chooses $M_0$ to be a projection onto a state orthogonal to $\ket{\tilde{\varphi}_2}$ (see Fig.~\ref{fig:smalldelta}).
Thus, the probabilities of $M_0$ occurring when Alice sent $\ket{\tilde{\varphi}_0}$, $\ket{\tilde{\varphi}_1}$, $\ket{\tilde{\varphi}_2}$, and $\ket{\tilde{\varphi}_3}$ are $\sin^2(2\delta')$, $\sin^2(\delta')$, $0$, and $\sin^2(\delta')$, respectively.
Here, we denote $\delta'=\delta/2$.
Using $\sin(x)=x$ for small $x$ and \myeqnref{eqn-QBER1}, the QBER is
$(\frac{1}{2} \delta'^2+\frac{1}{2}\delta'^2)/(6\delta'^2) =
\frac{1}{6} = 16.7\%$. Note that this value is just a little bit
greater than the QBER of $15.5\%$ of our optimal attack strategy
plotted in Fig.~\ref{fig:upperbound_BB84}. Also note that
$M_{\text{vac}}$ is equal to
$\ket{\tilde{\varphi}_2}\bra{\tilde{\varphi}_2}$ with a probability
of occurrence of $1-3\delta'^2/2$ (it is $3\delta'^2/2$ for $M_0$),
thereby introducing a discontinuity in QBER at $\delta=0$.

The significance of
Fig.~\ref{fig:upperbound_BB84}
is that there is a range of phase differences $\delta$ that causes the QBER to go below $20\%$, which is shown in Ref.~\cite{Chau2002,Ranade2006} to be a tolerable QBER in BB84 when Eve does not have the ability to change the $\delta$.
This proves that Eve's ability to change the phase difference between Alice's states is helpful to Eve in breaking the security of BB84.
Specifically, when Alice and Bob are unaware of our Trojan horse attack, Eve can learn some information on the final key shared by Alice and Bob.
This can be seen as follows:
Suppose Eve launches this attack and induces a QBER of, say, $15.6\%$.
Since this is lower than $20\%$ which is when the key distillation technique in Ref.~\cite{Gottesman2003} is applicable,
Alice and Bob decide to apply this technique to distill a final key.
On the other hand, the result of Ref.~\cite{Curty2004} says that no secret key can be established between Alice and Bob when Eve launches an intercept-and-resend attack.
Thus, the final key shared by Alice and Bob is not completely secret and Eve has some information on it.

It is important that the transmittance (which is the fraction of Alice's signals received by Bob) in the case of Eve launching this attack is similar to that when Eve is not present and the system is in normal operation, since, otherwise, Bob may be able to notice Eve's intervention by observing the unusually low transmittance.
Obviously, the quantum channel loss directly affects the transmittance.
In our intercept-and-resend attack, Eve can avoid her signals experiencing the quantum channel loss.
Specifically, she can perform her measurement at the output port of Alice, and send her measurement result classically to her ally located at Bob's side.
Her ally then resends a signal, based on the measurement result, to Bob.
In this way, no channel loss is experienced by Eve (assuming that the classical channel is perfect).
However, this does not mean that the transmittance in our attack is one.
This is because, based on the Eve's measurement result, she occasionally sends a vacuum state to Bob, thus reducing the transmittance.
In a typical experimental setup \cite{Gobby2004}, the loss in the fiber is about $0.2$ dB/km.
Thus, with an $100$ km-long fiber, the transmittance is about $10^{-\frac{0.2 \times 100}{10}}=0.01$.
In our intercept-and-resend attack that minimizes the QBER, it can be shown that for $\delta > \pi/20$, transmittance greater than $0.01$ can be achieved.
From Fig.~\ref{fig:upperbound_BB84}, when $\delta = \pi/20$, the QBER is about $15.6\%$.
This means that Eve can induce the same transmittance as in the normal operation of the system and still she can learn some information about the final key shared by Alice and Bob.

We remark that the POVM $\{M_{\text{vac}},M_i:i=0,\ldots,3\}$ of our
intercept-and-resend attack
is feasible
with current technology since each POVM element $M_i$ is a projection onto some state and
can be implemented as one direction of an orthogonal projection.
Thus, multiple orthogonal projections can be arranged to realize the projections of the POVM element $M_i$.

\subsection{Attack extensions}

We may further improve our attack by allowing Eve to send arbitrary
states to Bob with arbitrary number of POVM elements.
Note that changing the states sent to Bob only affects the penalty
values in the QBER (i.e., the three coefficients appearing before
the three states in \myeqnref{eqn-QBER2} are affected). By using a
similar analysis as in Ref.~\cite{Fung2006}, we obtain a QBER of
$14.6\%$ in this case, about $1\%$ lower than the case of Eve
sending BB84 states to Bob. The QBER upper bound with this
improvement is shown in Fig.~\ref{fig:upperbound_BB84} as the second
curve from the top.

\begin{figure}
\includegraphics[width=\figlength]{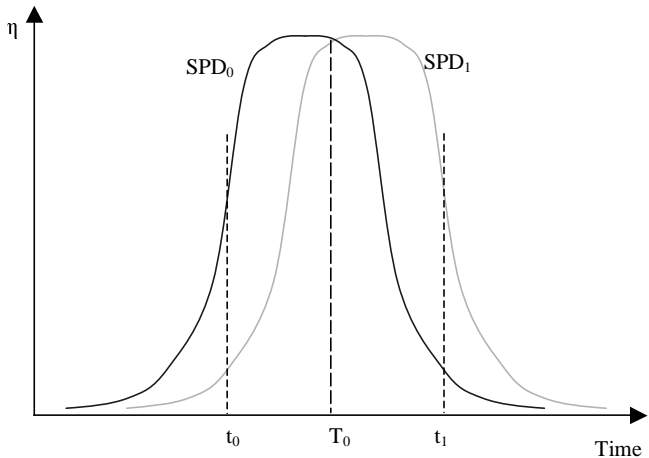}
\caption{\label{fig:spd} Efficiencies of two detectors. When Eve
time shifts the signals to arrive at Bob at time $t_0$ ($t_1$), the
efficiency of detector SPD0 (SPD1) is higher than that of detector
SPD1 (SPD0).}
\end{figure}
We may combine our phase-remapping attack with another Trojan horse
attack proposed in Ref.~\cite{Makarov2006}, a fake signals attack,
to obtain even further improvement on the QBER upper bound. In the
fake signals attack, Eve takes advantage of the detector efficiency
mismatch by time shifting the signals entering Bob's detector
package. Essentially, by time shifting the arriving signal from the
normal arrival time, the efficiency of the detector for detecting
``0'' becomes different from the efficiency of the detector for
detecting ``1'' (see Fig.~\ref{fig:spd}). Eve may make use of this
difference in the efficiencies to her advantage.
Ref.~\cite{Makarov2006} proposed a specific intercept-and-resend
attack with a fixed measurement (the normal BB84 measurement in the
$X$ and $Z$ bases) and fixed resent states (the normal BB84 states
but with the time shifted) and showed that it is possible to
compromise the QKD system if Alice and Bob are unaware of this
attack. Here, we combine our phase-remapping attack with the
fake signals attack. Specifically, Eve performs phase remapping of
Alice's states (which is also achieved by time shifting), measures
Alice's output signals, and resends to Bob some signals having the
arrival time shifted from the normal arrival time. We may proceed to
compute the QBER upper bound
 by minimizing the QBER over arbitrary POVM measurements but with the same resent states as those proposed in Ref.~\cite{Makarov2006}
 (e.g., when Eve detects the state
$\ket{\tilde{\varphi}_{0}}$, she resends the $\ket{-}$ state time
shifted to a location where the detector for bit ``0'' has a higher
efficiency). The QBER is the same as \myeqnref{eqn-QBER1} but with
different $L_i$ and $B_i$ for this attack.  For example, those
corresponding to sending the $\ket{-}$ state are
\begin{eqnarray}
\label{eqn-L0}
 L_0 &=& \frac{1}{2} \eta_1(t_0)
\ket{\tilde{\varphi}_{0}}\bra{\tilde{\varphi}_{0}} + \eta_1(t_0)
\ket{\tilde{\varphi}_{1}}\bra{\tilde{\varphi}_{1}} + \nonumber \\
&& \frac{1}{2} \eta_0(t_0)
\ket{\tilde{\varphi}_{2}}\bra{\tilde{\varphi}_{2}} \\
\label{eqn-B0}
 B_0 &=&
 \frac{1}{2} (\eta_0(t_0)+
 \eta_1(t_0)
 )
 \big[\ket{\tilde{\varphi}_{0}}\bra{\tilde{\varphi}_{0}}
 +\ket{\tilde{\varphi}_{2}}\bra{\tilde{\varphi}_{2}}\big] + \nonumber \\
&& \eta_1(t_0)
 \big[\ket{\tilde{\varphi}_{1}}\bra{\tilde{\varphi}_{1}}
 +\ket{\tilde{\varphi}_{3}}\bra{\tilde{\varphi}_{3}}\big],
\end{eqnarray}
where $\eta_0(t_0)$ ($\eta_1(t_0)$) is the efficiency of the
detector for bit ``0'' (``1'') at time $t_0$.  This combinational
attack results in the third curve from the top in
Fig.~\ref{fig:upperbound_BB84}, with the assumption that the
efficiency mismatch between the two detectors
(i.e., $\eta_1(t_0)/\eta_0(t_0)$) is $0.08$. Furthermore, by
minimizing the QBER over the measurements and also the states resent
by Eve, we obtain the bottom curve in
Fig.~\ref{fig:upperbound_BB84}, with the same efficiency mismatch.
As shown in the figure, there is considerable improvement in the
QBER upper bound by combining with the fake signals attack. Note
that the fake signals attack alone corresponds to the endpoints of
the bottom two curves at $\delta=\pi/2$ (the QBER values are
$12.3\%$ and $9.82\%$). Moving along the bottom curve, we see that
by combining with our phase-remapping attack, the QBER upper bound
decreases significantly from $9.82\%$ to $5.79\%$.

\begin{figure}
\includegraphics[width=\figlength]{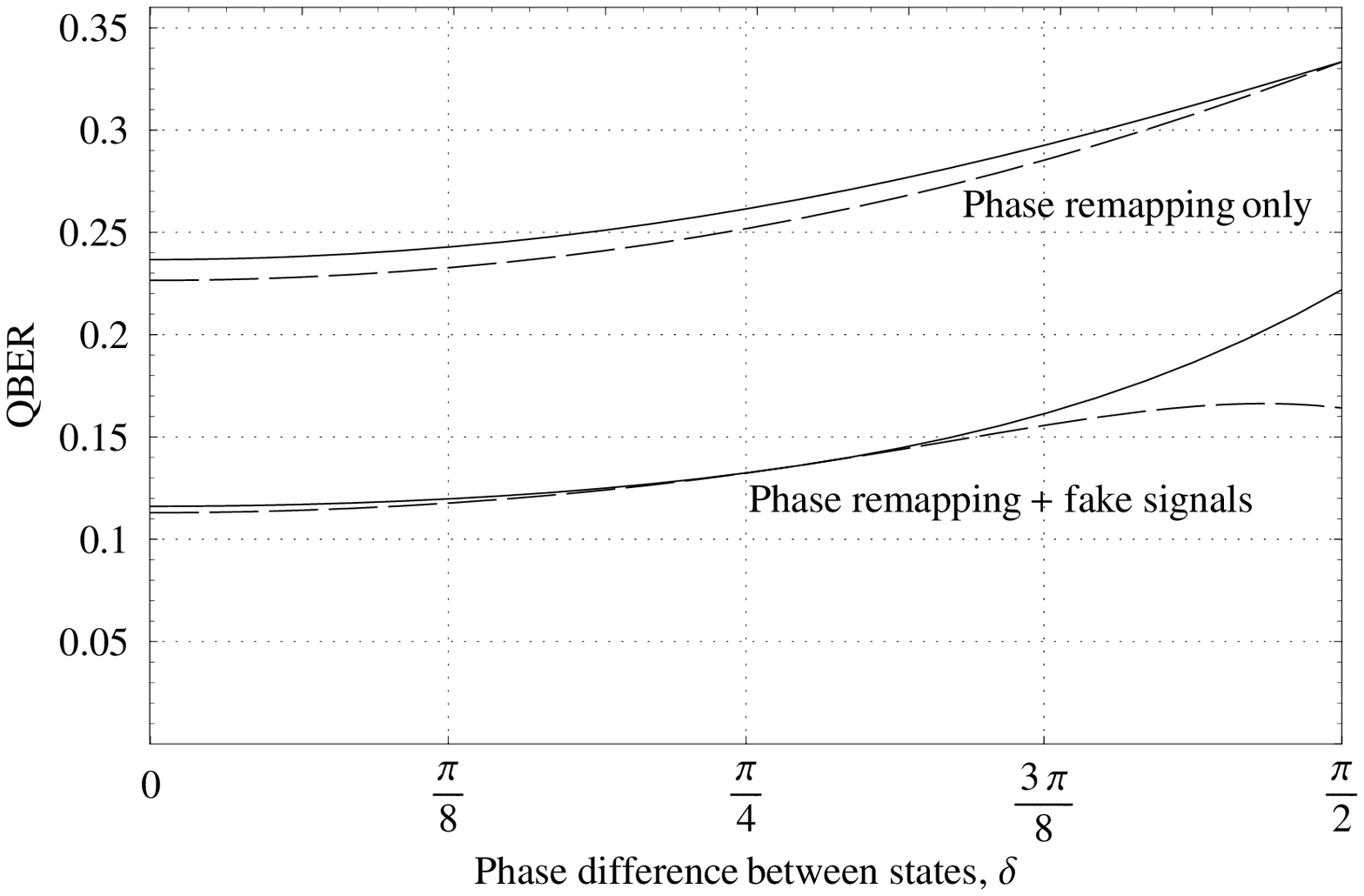}
\caption{\label{fig:upperbound_SARG04}
QBER upper bound of Trojan horse attacks for SARG04.
The top two curves correspond to the phase-remapping attack only whereas the bottom two cures correspond to the combination of the phase-remapping attack and the fake signals attack of Ref.~\cite{Makarov2006}.
The QBER of the two solid curves are obtained by
minimizing over the POVM measurement by Eve for each phase difference $\delta$ and assuming a fixed state sent to Bob.
The QBER of the two dashed curves are obtained by
minimizing over the POVM measurement by Eve and the state sent to Bob for each phase difference $\delta$.
Note that the QBER values approach some minimum values ($23.7\%$, $22.7\%$, $11.6\%$, and $11.3\%$) as the phase difference between the states approaches zero.
}
\end{figure}
Our phase-remapping attack and also the fake signals attack work against not only on the BB84 protocol,
but also the Scarani-Acin-Ribordy-Gisin 2004 (SARG04) protocol \cite{Scarani2004}.
We have plotted an analogous figure for
the SARG04 protocol
in Fig.~\ref{fig:upperbound_SARG04}.
The methods for obtaining these curves are similar to that for the BB84 protocol.
In this figure, we have also used the efficiency mismatch of $0.08$ for the fake signals attack.
We remark that the tolerable QBER for the SARG04 protocol is $19.9\%$ \cite{Fung2006} when Alice and Bob are not aware of any Trojan horse attacks.
Similar to the conclusion for the BB84 protocol,
since, as shown in Fig.~\ref{fig:upperbound_SARG04}, the QBER values induced by our phase-remapping attack together with the fake signals attack for a large range of phase difference $\delta$ are below the tolerable QBER,
the security of the SARG04 protocol can be compromised.


\section{Phase-remapping attack with a weak coherent-state source\label{sec-coherent}}

In this section, we consider the phase-remapping attack when a weak coherent-state source is used, which is in contrast to
Sec.~\ref{sec-singlephoton} where a single-photon source is assumed.
Here, we aim to show that there exist some situations where a normal post-processing would lead Alice and Bob to wrongly believe that the secret key generation rate is positive but in fact it is zero.
In order to ensure that no secret can be extracted, we again make Eve perform the time-shifting operation to achieve phase remapping followed by an intercept-and-resend attack as in Sec.~\ref{sec-singlephoton}.
This time, however, Eve may perform additional operations before her intercept-and-resend attack.
Since there can be more than one photon in a signal pulse traveling from Alice to Bob,
Eve may perform a quantum non-demolition (QND) measurement to determine the number of photons in the signal and
then an intercept-and-resend attack that may dependent on the photon number.
However, in the three strategies that we will discuss
below,
Eve does not need to perform such a QND measurement.
Indeed, our three strategies are feasible with current technology.
In any case, any entanglement carried by any signal from Alice to Bob is destroyed by Eve's attack, regardless of the number of photons in the signal,
since an intercept-and-resend attack corresponds to an entanglement-breaking channel.
Therefore, the secret key generation rate must be zero \cite{Curty2004}.


On Alice and Bob's side, we adopt a specific post-processing step
after the sifted key is obtained. Specifically, Alice and Bob
establish security using the result of
Gottesman-Lo-L{\"{u}}tkenhaus-Preskill (GLLP) \cite{Gottesman2004}
(which assumes the worst-case estimations for the proportion of the
single-photon signals and their QBER)
and they optionally perform
two-way classical post processing (using B steps
\cite{Gottesman2003}).
 However, the two-way post-processing step
in Ref.~\cite{Gottesman2003} cannot be applied directly, since a
single-photon source is assumed there, whereas we are considering a
weak coherent-state source here. Instead, we apply the two-way
post-processing technique for weak coherent-state sources proposed
by us in Ref.~\cite{Ma2006} (although decoy states are used there,
we will directly apply the technique without decoy states here).
Afterwards, they perform standard error correction and privacy amplification to distill the final key.
We summarize a QKD model for realistic setups, a key generation rate formula for a weak coherent-state source, and a two-way post-processing procedure using B steps for a weak coherent-state source in Appendix~\ref{app-QKDmodel}.
This background material will be used later in this section.

Let us construct three specific examples in which Eve can
successfully trick Alice and Bob into believing that a secret key
can be generated. We adopt a model in which all imperfections are
attributed to Eve (as in Ref.~\cite{Gottesman2004,Lo2005,Ma2005b})
or, viewed from a different perspective, Eve can control the quantum
channel and the detectors.
In both examples,
she treats all signals with two and more photons as single-photon signals and
performs an intercept-and-resend attack 
on all non-vacuum signals.
In the intercept-and-resend attack, we assume for simplicity
that Eve's measurement only identifies the states
$\ket{\tilde{\varphi}_0}$ and $\ket{\tilde{\varphi}_3}$ and resends
some arbitrary states to Bob
\footnote{In the plug \& play system, time shifting the signals
to be modulated can only decrease the phase difference $\delta$
between Alice's four states from the normal BB84 phase difference of
$\pi/2$. In this case, only resending for the detections of the
states $\ket{\tilde{\varphi}_0}$ and $\ket{\tilde{\varphi}_3}$
can induce a smaller QBER than the normal BB84 threshold of $25\%$.
Thus, we assume that Eve resends only when she detects the states
$\ket{\tilde{\varphi}_0}$ and $\ket{\tilde{\varphi}_3}$.
Alice and Bob can in principle monitor the statistics of the four
states and may notice the abnormality, which may lead them to think
that Eve may be interfering with the channel. On the other hand, for
the Sagnac system, the statistics of the four states can be made the
same as in normal BB84, since any phase difference (larger or
smaller than the normal BB84 phase difference) can be chosen by Eve.
Thus, examples with no abnormality in the statistics may be
constructed. Here, for simplicity of the analysis, we assume that
Eve resends only the two aforementioned states. }. The
intercept-and-resend attack is optimized for the phase difference
$\delta$ that Eve has chosen to remap Alice's four states.
Note that it is not difficult to construct intercept-and-resend
attacks specific to signals of certain numbers of photons in a
similar way as that for the single-photon signals. The first example
demonstrates the phase-remapping attack alone with a weak
coherent-state source. The second and third examples illustrate
mixed attack strategies that combine the phase-remapping attack and
the fake signals attack; and these two examples differ in whether
or not Eve fine tunes her attack strategy to match the overall gain
and the overall QBER (see Appendix~\ref{app-QKDmodel} for their
definitions) with the normal operating values.

\subsection{Strategy one}

In this strategy,
Eve performs phase remapping followed by intercepting Alice's signal
and resending only the states $\ket{0_z}$ and $\ket{1_x}$ (with
equal probabilities) to Bob \cite{endnote39}. This strategy produces
the following overall gain and overall QBER, respectively,
\begin{eqnarray}
Q_{\text{signal}} &=& p_{\text{dark}} e^{-\mu} + (C_1+(1-C_1)p_{\text{dark}}) \nonumber \\
\label{eqn-EveQsignal}
&& (1- e^{-\mu}) \\
E_{\text{signal}} &=&  \big[p_{\text{dark}} e^{-\mu}/2 + (C_1 e_1 +(1-C_1)p_{\text{dark}}/2) \nonumber \\
\label{eqn-EveEsignal}
&& (1- e^{-\mu}) \big] /Q_{\text{signal}},
\end{eqnarray}
where $p_{\text{dark}}$ is the dark count probability, $e_1$ and $C_1$ are, respectively, the QBER and the conclusive probability of the intercept-and-resend attack for the single-photon case.
If there is no detection error (i.e. $e_{\text{detector}}=0$ and it is the case in this example),
$e_1$ can be computed from Eqs.~\eqref{eqn-QBER1}-\eqref{eqn-QBER3}
or extracted from the top curve of Fig.~\ref{fig:upperbound_BB84}
for a particular phase difference $\delta$ (since the top curve
of Fig.~\ref{fig:upperbound_BB84} is achieved by Eve resending only
the states $\ket{0_z}$ and/or $\ket{1_x}$ to Bob). On the other
hand, if $e_{\text{detector}}$ is not zero, we need to incorporate
it in the calculation of $e_1$, which can be easily done.

Note that both states $\ket{0_z}$ and $\ket{1_x}$ sent by Eve to Bob cause the same QBER's and the same gains on Bob's side
(since their phase-remapped states $\ket{\tilde{\varphi}_0}$ and $\ket{\tilde{\varphi}_3}$ in \myeqnref{eqn-phaseremappedstates} are symmetrical (see Fig.~\ref{fig:smalldelta})).
The conclusive probability $C_1$, which is also the probability that
Eve resends the states $\ket{0_z}$ and $\ket{1_x}$, is equal to
$C_1=\operatorname{Tr}((M_0+M_3)B)/4$,
where $M_0$ and $M_3$, with $M_0+M_3\leq I$, are the POVM elements
for resending the two states obtained by minimizing the QBER $e_1$,
and $B$ as given in \myeqnref{eqn-QBER3} is the density matrix sent
by Alice to Eve.
Also, we assume that Eve always sends a strong pulse to Bob, which
is reflected in the exclusion of Bob's detector efficiency
$\eta_{\text{Bob}}$ in
Eqs.~\eqref{eqn-EveQsignal}-\eqref{eqn-EveEsignal}, $C_1$, and
$e_1$.


\begin{table}
\centering
\begin{tabular}{|c|c|c|c|c|}
\hline
$\alpha$ [dB/km] & $\eta_{\text{Bob}}$ & $e_{\text{detector}}$ & $p_{\text{dark}}$ & $f$\\ \hline
$0.21$ & $8.0\%$ & $0\%$ & $10^{-7}$ & $1.16$\\ \hline
\end{tabular}
\caption{\label{table-sim-param}
Simulation parameters.
Here,
$\alpha$ is the channel loss coefficient,
$\eta_{\text{Bob}}$ is the detector efficiency,
$e_{\text{detector}}$ is the detection error probability,
$p_{\text{dark}}$ is the dark count rate,
and $f$ is the error correction inefficiency.
See Appendix~\ref{app-QKDmodel} for detail.
}
\end{table}
Since Alice and Bob use only the result of GLLP to ensure security, the mean photon number $\mu$ they use may be very small.
(In contrast, when the decoy-state method \cite{Hwang2003,Lo2004,Lo2005,Wang2005a,Wang2005b,Ma2005b,Harrington2005} is used to ensure security, the mean photon number may be high, e.g. on the order of $1$.)
Suppose that the mean photon number is $\mu=8\times10^{-4}$ and three B steps are used by Alice and Bob.
\begin{figure}
\includegraphics[width=\figlength]{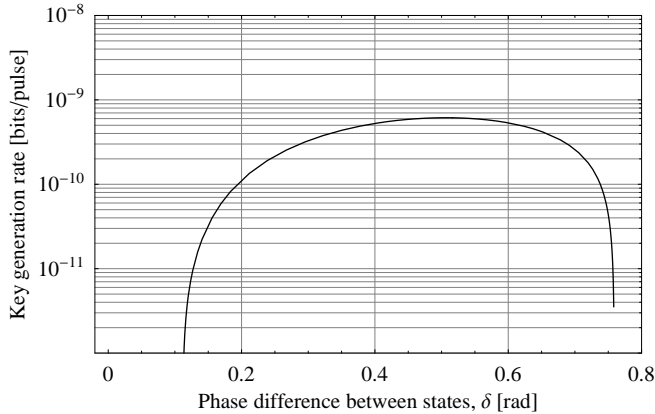}
\caption{\label{fig:coherent_rateangle}
Key generation rates at various distances.
We use the QKD model parameters shown in Table~\ref{table-sim-param} to compute the overall gain and the overall QBER from Eqs.~\eqref{eqn-EveQsignal}-\eqref{eqn-EveEsignal}.
The key generation rates are then computed using \myeqnref{eqn-GLLPKeyRate} with three B steps for various distances.
Here, the key rates should be zero (since Eve launches an intercept-and-resend attack) but are positive in the range $0.12 \leq \delta \leq 0.75$, meaning that the keys generated in this range are insecure.
}
\end{figure}
We use the QKD model parameters shown in Table~\ref{table-sim-param} to compute the overall gain and the overall QBER from Eqs.~\eqref{eqn-EveQsignal}-\eqref{eqn-EveEsignal}.
We can then compute the key generation rates using \myeqnref{eqn-GLLPKeyRate} for various distances, as shown in Fig.~\ref{fig:coherent_rateangle}.
The important point is that there is a range of phase differences ($0.12 \leq \delta \leq 0.75$) where the key generation rates are positive, but in fact no key can be generated since Eve's intercept-and-resend attack corresponds to an entanglement-breaking channel \cite{Curty2004}.
This means that the final keys generated in this range are insecure.
The key generation rates outside this range is zero with this particular strategy.
In contrast to this strategy, the two strategies that we describe next combine the phase-remapping attack with the fake signals attack \cite{Makarov2006}.

\subsection{Strategy two}
This strategy combines the phase-remapping attack with the fake
signals attack \cite{Makarov2006}. Specifically, in this strategy,
Eve performs phase-remapping followed by intercepting Alice's
signals and resending a time-shifted single-photon signal of
arbitrary state to Bob. Note that one crucial difference
between this strategy and strategy one is that here Eve takes
advantage of the efficiency mismatch of the detectors by time
shifting her signals sent to Bob. To simplify the analysis, we
assume that Eve always sends single-photon signals to Bob (in which
case the ratio of the efficiencies is the largest (cf.
\myeqnref{eqn-eta-n} and double clicks due to multiple photons of
arbitrary states are avoided). We compute the overall gain and the
overall QBER by using \myeqnref{eqn-EveQsignal} and
\myeqnref{eqn-EveEsignal} respectively. Here, we also assume that
Eve only resends when she detects $\ket{\tilde{\varphi}_{0}}$ and
$\ket{\tilde{\varphi}_{3}}$ (as in strategy one); and thus the
resending probability is $C_1=\operatorname{Tr}(M_0 B_0+M_3 B_3)/4$
where $B_i$ is from \myeqnref{eqn-B0}. We allow Eve to resend
arbitrary states to Bob; thus $e_1$ can be extracted from the bottom
curve of Fig.~\ref{fig:upperbound_BB84} for a particular phase
difference $\delta$ (if the efficiency mismatch is $0.08$) or
computed from Eqs.~\eqref{eqn-QBER1}, \eqref{eqn-L0},
\eqref{eqn-B0}, and the corresponding equations for $L_3$ and $B_3$.

\begin{table}
\centering
\begin{tabular}{|c|c|c|}
\hline
 $\eta_0(t_0)/\eta_1(t_0)$ &
 $\delta$ &

 Key rate \\
 \hline
 $0.0667$ & $1.02$ & $8.921 \times 10^{-7}$ ($2.610\times 10^{-7}$)  \\
 \hline
 $0.04$ & $1.31$ & $1.622 \times 10^{-6}$ ($1.457\times 10^{-6}$)\\
 \hline
 $0.03$ & $1.41$  & $2.038 \times 10^{-6}$ ($1.968\times 10^{-6}$)\\
 \hline

\end{tabular}
%
%
\caption{\label{table-PRTS1}
 Key generation rates for strategy two, in which Eve combines the
 phase-remapping attack with a fake signals attack.
 The first column is the efficiency mismatch of the two detectors (related to the fake signals attack);
 the second column is the phase difference between the states sent by
 Alice chosen to maximize the key generation rate
 (related to the phase-remapping attack);
the third 
 column is the key generation rate for the phase difference in the second 
 column.
The rates in
brackets correspond to the case of only the fake signals attack
without the phase-remapping attack. Note that there is some
improvement in the key rates by combining both attacks. We used a
mean photon number of $\mu=8\times10^{-4}$.
 }
\end{table}

We assume Alice and Bob use only the result of GLLP to ensure
security and no B step is used. We use the QKD model parameters
shown in Table~\ref{table-sim-param} and  $\mu=8\times10^{-4}$ to
compute the overall gain and the overall QBER for this strategy. We
then compute the key generation rates using
\myeqnref{eqn-GLLPKeyRate} for a few cases, and the result is
tabulated in Table.~\ref{table-PRTS1}. Here, we assume that when Eve
detects $\ket{\tilde{\varphi}_{0}}$ ($\ket{\tilde{\varphi}_{3}}$),
she time shifts the signal to arrive at Bob at time $t_0$ ($t_1$) as
in Fig.~\ref{fig:spd} and we assume symmetry between the two
detectors such that $\eta_0(t_0) = \eta_1(t_1)=\eta_\text{Bob}$ and
$\eta_1(t_0) = \eta_0(t_1)$, where $\eta_i(t)$ is the efficiency of
detector $i$ at time $t$. As shown in the table, the key generation
rates are positive but should be zero since this strategy is an
intercept-and-resend attack strategy \cite{Curty2004}. Therefore,
the final key Alice and Bob distill is compromised by Eve.  Note
that the key generation rates of this strategy are higher than that
of strategy one. One drawback of this strategy is that the overall
gain
 and the overall QBER
 induced by Eve may be quite different from what Alice and Bob may expect in a normal situation.
 To overcome this, we discuss a third strategy below that matches the induced gain and QBER with the normal operating values.
 Nevertheless, with this example, we have demonstrated that our phase-remapping attack in combination with the fake signals attack can compromise the security of the QKD system if Alice and Bob are unaware of the attack strategy.

\subsection{Strategy three} In this strategy, Eve also performs a
combination of the phase-remapping attack and the fake signals
attack \cite{Makarov2006} as in strategy two, but here she adjusts the parameters of
her attack to match the overall gain and the overall QBER with what
Alice and Bob would expect in normal cases. Alice and Bob may have
some idea on the parameters of their system and may have certain
expectation on the overall gain and QBER. Thus, Eve needs to adjust
her attack in order to simulate a normal situation. She does this by
altering the dark count probability of Bob's detectors (as stated
before, we assume that the detectors are under Eve's control) and
changing the resending probability in the intercept-and-resend
attack. Other than these two adjustments, strategy three is
otherwise the same as strategy two. In this strategy, the overall
gain and overall QBER are, respectively,
\begin{eqnarray}
\label{eqn-EveQsignal2}
Q_{\text{signal}} &=& Y_0 e^{-\mu} + (\gamma C_1+(1-\gamma C_1)Y_0) (1- e^{-\mu}) \\
E_{\text{signal}} &=&  \big[Y_0 e^{-\mu}/2 + (\gamma C_1 e_1 +(1-\gamma C_1)Y_0/2) \nonumber \\
\label{eqn-EveEsignal2}
&& (1- e^{-\mu}) \big] /Q_{\text{signal}},
\end{eqnarray}
where $Y_0$ is the dark count probability Eve chooses (which can be
different from the normal dark count probability $p_{\text{dark}}$)
and $0\leq\gamma\leq1$ is the resending probability for conclusive
results. The other variables are the same as in
strategy two.

We assume that the normal situation is produced by the QKD model
parameters shown in Table~\ref{table-sim-param} and
$\mu=8\times10^{-4}$.
From these parameters, the normal operating values of the overall QBER and the overall gain can be computed from Eqs.~\eqref{eqn-normalQsignal}-\eqref{eqn-normalEsignal}.
Eve then chooses the phase difference $\delta$, the dark count probability $Y_0$,
and the resending probability $\gamma$
for a fixed efficiency mismatch to match the overall QBER induced by
her (\myeqnref{eqn-EveQsignal}) and the overall gain induced by her
(\myeqnref{eqn-EveEsignal}) within $10\%$ of the normal operating
values.
We assume that Eve does not interfere with the detection error
probability; thus, we still have $e_{\text{detector}}=0$ as in the
normal situation and the QBER of the single-photon signals, $e_1$,
is computed as in strategy two. We show in
Table~\ref{table-show-attack1} two instances in which Eve's
combination of the phase-remapping attack and the fake signals
attack
achieves positive key generation rates.
\begin{table}
\centering
\begin{tabular}{|c|c|c|c|c|c|}
\hline
Distance $[$km$]$& $\frac{\eta_0(t_0)}{\eta_1(t_0)}$ & $\delta$ & $Y_0$ & $\gamma$ & Key rate\\
\hline

$88.0$ & $0.04$ & $1.31$ & $1\times 10^{-9}$ & $0.096$ & $4.057
\times 10^{-8}$ \\ \hline

$87.0$ & $0.03$ & $1.41$ & $1.8\times 10^{-8}$ & $0.1$ & $5.838
\times 10^{-8}$ \\ \hline

\end{tabular}
\caption{\label{table-show-attack1} Two situations in which Eve's
attack produces the same overall gain and overall QBER as that
produced in a normal situation described by the parameters in
Table~\ref{table-sim-param}. Here, Eve fine tunes her attack by
adjusting the phase difference $\delta$, the dark count probability
$Y_0$, and the resending probability $\gamma$ for some distance and
some efficiency mismatch between the two detectors.
We assume that Alice and Bob perform the post-processing steps from
GLLP and no B step as described in Appendix~\ref{app-QKDmodel}. The
fact that the key generation rates, computed using
\myeqnref{eqn-GLLPKeyRate},
are positive means that Eve has successfully compromised the final
keys. We used a mean photon number of $\mu=8\times10^{-4}$. }
\end{table}
In both instances, Alice and Bob simply use the post-processing
steps from GLLP and no B step to distill secret keys as described
earlier, with the QKD model parameters shown in
Table~\ref{table-sim-param} and  $\mu=8\times10^{-4}$.
In this example, both the normal situation and the hostile situation look similar to Alice and Bob.
The normal situation arises when Eve is not present while the hostile situation arises when
Eve launches this attack strategy.
Since both situations give rise to the same overall QBER and overall
gain, Alice and Bob are unaware of which situation they are in and
thus distill keys at the same key generation rate in both
situations. However, no secret key can be generated in the hostile
case, since it corresponds to an entanglement-breaking channel
\cite{Curty2004}. Thus, if Alice and Bob are unaware of the Trojan
horse attack, they may generate keys that are compromised by Eve.

Note that the values of the dark count probability $Y_0$ in
Table~\ref{table-show-attack1} are lower than the normal value given
in Table~\ref{table-sim-param}. While lowering the dark count
probability may be difficult to achieve in practice, Eve may realize
this strategy by increasing the dark count probability in the normal
situation instead.  In addition, we point out that dark count
probability on the order of $10^{-9}$ has been attained
experimentally \cite{Takesue2006}; thus, the values of the dark
count probability $Y_0$ shown in Table~\ref{table-show-attack1} are
realistic. We also note that the discontinuity in
Fig.~\ref{fig:upperbound_BB84} at $\delta=0$ does not manifest as a
problem in this attack for the weak coherent-state source.
This is because the phase difference $\delta$ is chosen to 
match
the overall QBER and gain
with some normal operating values.
In normal scenarios, 
$\delta$ is set to some non-zero value.

We remark that although the key generation rates in the three
examples may not be very significant,
they do raise
the awareness that the Trojan horse attack we propose
can be detrimental to Alice and Bob.


\section{Conclusions\label{sec-conclusions}}
We have proposed a realistic Trojan horse attack, the
phase-remapping attack,  for two-way quantum key distribution
systems implementing the BB84 protocol. We have shown that, when
Alice and Bob are unaware of our attack, there are situations
in both the perfect-single-photon-source case and the
weak-coherent-state-source case that the final key shared between
them is compromised and Eve has some information on it.
Specifically, for the perfect-single-photon-source case, when the
QBER is larger than $14.6\%$, Alice and Bob may distill a
compromised key. For the weak-coherent-state-source case, we have
given three examples (two of which are combined with a fake signals
attack) in which the final keys are insecure. Note that our attack
is feasible with only current technology and thus is highly
practical for Eve to implement. Therefore, it is important for Alice
and Bob to be aware of the possibility of our attack and to guard
against it by only generating a key when the QBER is low enough.

We remark that the fact that we demonstrated the insecurity of a key
guaranteed to be secure by some existing security proofs does not
imply that the proofs are incorrect. It is because the Trojan horse
attack we demonstrated corresponds to performing operations and
using information lying outside the Hilbert space assumed in the
proofs. These extra operations and information are granted to us by
the practical implementations of the BB84 protocol.  Thus, while a
QKD protocol may be unconditionally secure, a realistic
implementation of it may open up security loopholes via extra
dimensions.

\appendix

\section{Minimization of QBER\label{app-minQBER}}
The normalized bit error rate is (c.f. \myeqnref{eqn-QBER1})
\begin{eqnarray}
\text{QBER} &=& \frac{ \sum_{i=0}^{3} \sum_{j=0}^{1} \bra{j_z}W_i
L_i W_i^\dagger \ket{j_z} }{ \sum_{i=0}^{3} \sum_{j=0}^{1}
\bra{j_z}W_i B_i W_i^\dagger \ket{j_z} },
\end{eqnarray}
where $L_i$ and $B_i$ are given in \myeqnref{eqn-QBER2} and
\myeqnref{eqn-QBER3}, respectively, and $W_i^\dagger W_i \triangleq
M_i$ are the POVM elements. We want to minimize $\text{QBER}$ over
the eight independent
row
vectors $\bra{j_z}W_i$ each with two elements. At least
one of the eight must be non-zero, because otherwise all $W_i$ would
be zero and there would be no qubits sent to Bob. Since
$\text{QBER}$ is not a sum of eight independent ratios, i.e.,
\begin{eqnarray}
\text{QBER} &\neq& \sum_{i=0}^{3} \sum_{j=0}^{1}
\frac{
 \bra{j_z}W_i L_i W_i^\dagger \ket{j_z}
}{
 \bra{j_z}W_i B_i W_i^\dagger \ket{j_z}
},
\end{eqnarray}
it may appear at first sight that the minimization of $\text{QBER}$ is not trivial.
However, it turns out that we can minimize each ratio independently and set $\text{QBER}$ to be the smallest ratio by assigning zeros to the other seven vectors.
We show this by the following claim:
\noindent\begin{claim}
Given two ratios, $\frac{a_1}{a_2}$ and $\frac{b_1}{b_2}$,
if $\frac{a_1}{a_2} \leq \frac{b_1}{b_2}$,
then $\frac{a_1}{a_2} \leq \frac{a_1+b_1}{a_2+b_2}$.
\end{claim}
Therefore, we consider separately minimizing each ratio, which can be written as
\begin{eqnarray}
\label{eqn-app-RayleighQuotient}
\frac{
\langle c_{ji}| B_i^{-\frac{1}{2}} L_i B_i^{-\frac{1}{2}}
|c_{ji}\rangle }{
\langle c_{ji}|c_{ji}\rangle
},
\end{eqnarray}
where $\langle c_{ji}|=\bra{j_z}W_i B_i^{\frac{1}{2}}$ is a
row
vector with two elements.
The eigenvector of
$B_i^{-\frac{1}{2}} L_i B_i^{-\frac{1}{2}}$ corresponding to the
minimum eigenvalue minimizes \myeqnref{eqn-app-RayleighQuotient}.
The minimum eigenvalue among all $i$'s is the
minimum QBER, which is the top curve plotted in
Fig.~\ref{fig:upperbound_BB84}. It is not difficult to ensure that
the POVM elements satisfy $\sum_{i=0}^3 W_i^\dagger W_i \leq
\mathbf{I}$.
Note that we can always scale the POVM elements (by the same factor) without affecting the QBER.
Thus, it is always possible to find a scaling such that these POVM elements and an additional one corresponding to sending a vacuum state to Bob add up to identity.

\section{Review of QKD model and key generation rate for realistic setups\label{app-QKDmodel}}

We first review a widely-used model for realistic QKD setup (see, e.g., \cite{Lutkenhaus2000,Lo2005}).
This model is suitable for fiber-based QKD systems.
We then summarize the key generation rate from GLLP \cite{Gottesman2004} and the B step
\cite{Gottesman2003,Ma2006,Khalique2006}, for the weak-coherent-state-source case.

{\bf Source:} The source is a single-mode laser source.
We assume that the phase of each pulse is randomized.
Thus, the laser source
emits pulses that are a classical mixtures
of the photon number states with a Poisson distribution:
\begin{eqnarray}
\sum_{i=0}^{\infty} \frac{\mu}{i!}e^{-\mu} \ket{i}\bra{i},
\end{eqnarray}
where $\mu$ is the mean photon number.

{\bf Transmission:}
The quantum channel is the optical fiber and
we quantify the loss in the optical fiber by the probability that an input photon is lost at the end of the transmission.
Let $\alpha$ in dB/km be the loss coefficient of the optical fiber and $l$ be the fiber length in km.
Then, the probability that the input photon is not lost is equal to $10^{-\frac{\alpha l}{10}}$.

{\bf Detection:}
We assume Bob is equipped with threshold detectors.
Since they are not completely efficient, there is some chance that they do not produce a click even when there are some photons present at the inputs.
The probability that Bob's detector detects the presence of an input photon is defined as Bob's detection efficiency $\eta_{\text{Bob}}$.
%
Combining the loss in the quantum channel and the inefficiency of Bob's detector,
we arrive at the overall transmission efficiency, $\eta$.
It is the probability that a photon is detected given that one has been sent, and is given by
\begin{eqnarray}
\label{eqn-eta}
\eta &=& 10^{-\frac{\alpha l}{10}}  \eta_{\text{Bob}} .
\end{eqnarray}
When the input signal contains more than one photons, the signal is detected if at least one photon is detected.
Thus, the transmission efficiency for an $n$-photon signal is
\begin{eqnarray}
\label{eqn-eta-n}
\eta_n &=& 1-(1-\eta)^n.
\end{eqnarray}

When there is no input to Bob's detector, there is a possibility that it generates a detection event.
This is due to the intrinsic detector's dark counts, the background spray, and the leakage from timing signals.
We denote the probability of this false detection event as $p_{\text{detector}}$.
Suppose that there are two detectors in the system.
We denote the probability of false detection for the system as $p_{\text{dark}}=2p_{\text{detector}}(1-p_{\text{detector}})$.

When there is a double-click event, which occurs because of dark counts or detection of a multi-photon signal,
we impose that Bob takes one of the bit values randomly
\cite{Inamori2005,Gottesman2004}. This is consistent with the
so-called ``squash operation'' used in the security proof of
GLLP~\cite{Gottesman2004}.

More concretely, the security proof of GLLP assumes that the squash
operation is performed by Eve. This operation is a mapping from a
multi-photon state to a qubit state.
Thus, under this assumption, Eve always sends a qubit state to Bob.
In this paper, we directly apply the result of GLLP to our
calculations of key generation rates and therefore we assume the
squash operation without proof. We consider two-way classical
post-processing in this paper and our squash-operation assumption
simplifies our analysis.
We remark that Koashi \cite{Koashi2006} has proved the security of
one-way classical post-processing type QKD for a threshold detector
model without requiring the squash-operation assumption.

{\bf Yield, QBER, gain:}
Let us define the yield $Y_n$, the quantum bit error rate (QBER) $e_n$, and the gain $Q_n$.
The yield, $Y_n$, is defined as the probability that
Bob detects a signal conditional on Alice's $n$-photon emission:
\begin{eqnarray}
\label{eqn-def-yield}
Y_n &\triangleq&
Pr\{ \text{Detection by Bob} | \nonumber \\
&& \phantom{xxxxx} \text{Alice sent $n$-photon state} \} .
\end{eqnarray}
The yield is basically a sum of the probabilities of the error events and the no-error events.
The fraction of the error events in the total probability is the
quantum bit error rate $e_n$:
\begin{eqnarray}
\label{eqn-def-ber}
e_n &\triangleq&
Pr\{ \text{Bob's result is incorrect} | \nonumber \\
&&
\phantom{xxxxx}
\text{detection by Bob} \wedge \nonumber \\
&&
\phantom{xxxxx}
\text{Alice sent $n$-photon state} \} .
\end{eqnarray}
The gain of the $n$-photon state is
\begin{eqnarray}
\label{eqn-def-Qn}
Q_n &\triangleq&
Pr\{ \text{Detection by Bob} \wedge \nonumber \\
&&
\phantom{xxxxx}
\text{Alice sent $n$-photon state} \} \\
&=& Y_n e^{-\mu} \mu ^n / n! .
\end{eqnarray}
The overall gain and the overall QBER are the weighted averages of all the $n$-photon gains and QBER's:
\begin{eqnarray}
Q_{\text{signal}} &=& \sum_{n=0}^{\infty} Y_n e^{-\mu} \mu ^n / n! \\
E_{\text{signal}} &=& \frac{1}{Q_{\text{signal}}} \sum_{n=0}^{\infty} e_n Y_n e^{-\mu} \mu ^n / n! .
\end{eqnarray}
These two are parameters that Alice and Bob measure during a QKD experiment and can be used to determine the key generation rate \cite{Gottesman2004}.

{\bf Normal situation:}
When Eve is not present,
we assume that signals are emitted by the weak coherent-state source at Alice's side, travel through the optical fiber suffering some loss, and reach Bob on his detectors.
Under this situation, the normal values for the yields and the QBER for BB84 can be obtained as
\begin{eqnarray}
Y_n &=& p_{\text{dark}} (1-\eta_n) + \eta_n \\
e_n &=& \big[p_{\text{dark}} (1-\eta_n)/2 + \eta_n e_{\text{detector}} \big]/Y_n,
\end{eqnarray}
where $e_{\text{detector}}$ is a parameter representing the misalignment of the detector setup.
For the overall gain and the overall QBER, their normal values are
\begin{eqnarray}
\label{eqn-normalQsignal}
Q_{\text{signal}} &=& p_{\text{dark}} e^{-\mu \eta} + 1 - e^{-\mu \eta} \\
\label{eqn-normalEsignal}
E_{\text{signal}} &=& \frac{1}{Q_{\text{signal}}} \big[ \frac{p_{\text{dark}} e^{-\mu \eta}}{2}  + \nonumber \\
&&\phantom{xxxxxx}(1 - e^{-\mu \eta})e_{\text{detector}} \big].
\end{eqnarray}

{\bf Key generation rate:}
Once Alice and Bob have measured the overall gain and the overall QBER, the key generation rate may be obtained by using a result in GLLP~\cite{Gottesman2004} as follows:
\begin{eqnarray}
\label{eqn-GLLPKeyRate}
R &=& \frac{1}{2} r_B Q_{\text{signal}} \big[ -f(E_{\text{signal}}) H_2(E_{\text{signal}}) + \nonumber \\
&&\phantom{xxx}\Omega (1-H_2(e_p)) \big],
\end{eqnarray}
where $f(\cdot)$ is the error correction efficiency as a function of the QBER,
$H_2(p)=-p \log_2(p) - (1-p) \log_2(1-p)$ is the binary entropy function, $\Omega=Q_1/Q_{\text{signal}}$ is the fraction of single-photon states, $e_p$ is the phase error rate of the single-photon states,
and $r_B$ is the fraction of bits retained after B steps ($r_B=1$ if no B step is performed).
The factor of $1/2$ is the fraction of bits retained after basis reconciliation for BB84.
The first term in the bracket is related to error correction, while the second term is related to privacy amplification.
In this equation, 
$Q_1$ and $e_p$ are not directly measured, but they may be bounded by assuming the worst-case situation \cite{Gottesman2004}.
%
We may pessimistically assume that the overall gain $Q_{\text{signal}}$ is contributed by multi-photon signals as much as possible, and
all the errors come from single-photon detection events, leading to
$Q_1=Q_{\text{signal}}-p_{\text{multi}}$ and $e_1=E_{\text{signal}} Q_{\text{signal}}/Q_1$,
where $p_{\text{multi}}$ is the probability of Alice emitting multi-photon signals.
Before the post-processing using B steps (which we describe next), the phase error rate is equal to the bit error rate for the single-photon states, i.e. $e_p=e_1$.

{\bf B step:}
Optionally, Alice and Bob may perform one or more B steps by using two-way classical communications to increase the achievable secure distance.
The B step was analyzed in Ref.~\cite{Gottesman2003} for the single-photon source and in Ref.~\cite{Ma2006,Khalique2006} for the weak coherent-state source.
Each B step involves the following operations:
Alice and Bob first randomly pair up their bits, say $x_1$, $x_2$ on Alice's side and the corresponding $y_1$, $y_2$ on Bob's side.
They compute the parities of the pairs, $x_1 \oplus x_2$ and $y_1 \oplus y_2$, and publicly compare them.
If both parities are the same, they keep $x_1$ and $y_1$ and discard $x_2$ and $y_2$;
otherwise, they discard $x_1$, $x_2$, $y_1$, and $y_2$.
After each B step, the bit and phase error rates and the fraction of the single-photon states change.
We summarize the update formulas for the changes after running one B step as follows \cite{Ma2006}:
\begin{eqnarray}
\Omega' &=& \frac{\Omega^2 (e_1^2 + (1-e_1)^2)}{E_{\text{signal}}^2 + (1-E_{\text{signal}})^2} \\
E_{\text{signal}}' &=& \frac{E_{\text{signal}}^2}{E_{\text{signal}}^2 + (1-E_{\text{signal}})^2} \\
e_p' &=& \frac{2 e_p (1-e_1-e_p)}{e_1^2 + (1-e_1)^2} \\
e_1' &=& \frac{e1^2}{e_1^2 + (1-e_1)^2} \\
r_B' &=& r_B (E_{\text{signal}}^2 + (1-E_{\text{signal}})^2)/2,
\end{eqnarray}
where the primed (unprimed) variables are the new (old) values.
After running some number of B steps, we may obtain the key generation rate by using \myeqnref{eqn-GLLPKeyRate}.

\bibliographystyle{apsrev.bst}
\bibliography{paperdb}

\end{document}